# Stakeholder identification for a structured release planning approach in the automotive domain

Kristina Marner[1], Stefan Wagner[2], and Guenther Ruhe[3]

[1]Institute of Software Technology, University of Stuttgart, Stuttgart, Germany
Dr. Ing.h.c.F. Porsche AG, Weissach, Germany
kristina.marner@iste.uni-stuttgart.de

[2]University of Stuttgart, Universitätsstr.38,70569 Stuttgart, Germany
stefan.wagner@iste.uni-stuttgart.de

[3]Software Engineering Decision Support Laboratory, University of Calgary, Calgary AB, Canada
ruhe@ucalgary.ca

**Abstract.** Context: In regulated domains like automotive, release planning is a complex process. The agreement between traditional product development processes for hardware as well as mechanic systems and agile development approaches for software development is a major challenge. Especially the creation and synchronization of a release plan is challenging. Objective: The aim of this work is to present identified stakeholders of a release plan as an appropriate approach to create transparency in release planning in the automotive domain. Method: Action research to elaborate relevant stakeholders for release planning was conducted at Dr. Ing. h. c. F. Porsche AG. Results: We present a detailed overview of identified stakeholders due to release planning as well as their required content and added value regarding to two pilot projects. The results confirm the fact that almost every stakeholder is involved in a release plan in a certain way. Conclusions: Release planning within a complex project environment and complicated customer constellations is difficult to manage. We discuss how the presented stakeholders could meet with the given conditions in the automotive domain. With this contribution, identified stakeholders of release planning from hardware and software point of view is introduced. It helps to reach transparency and to handle the given complexity.

## 1 Introduction

A long time ago, vehicles were a purely mechanical matter. Current challenges in the automotive industry are connectivity, artificial intelligence or electric mobile services driven by environmental legislation [1]. Nowadays such a variety of software necessitates a corresponding planning and coordination effort. State-of-the-art system architecture of vehicles consists of up to 100 electronic control units (ECUs) distributed in the vehicle. However, the trend is towards centralization of functionality in less ECUs [2]. The centralization is a paradigm shift towards a more function-oriented development. Nevertheless, today's vehicles consist of an interaction between mechatronic systems, software and hardware [3]. The new view, the centralization of functions and these interactions cause serious challenges [4]. Furthermore, such complex development projects have to be considered and planned in a multidisciplinary way. The rising complexity because of the increasing embedded IT in vehicles make a common development more difficult considering all incorporated dependencies.

Release planning is a key activity for effective and efficient product implementation [5]. It describes the selection of an optimal subset of requirements that will be implemented in a particular release [6]. This coordination and control is based on a release plan that is essential for projects achieving the goals assigned to them. A major part of release planning is to consolidate complexity such as different stakeholders' and customers' needs as well as to fulfil boundary conditions given by the company into one release plan. The necessity for release planning is based on the fact that failures made at the beginning of a development are hardly compensated later during the development process. The creation and update of a common release plan of vehicles that includes all stakeholders is a central challenging task. This task is the more complex, the more people involved work according to different development processes [5].

The multidisciplinary work mentioned above causes a variety of development processes ranging from completely traditional, waterfall-like processes to adaptations of agile processes. Combining them in the development of a complete vehicle is necessary but hard to achieve. In particular, traditional development is milestone-driven, focused on high safety requirements and complex supplier constellations while agile development methods are applied to the development of control units, software and functions with regard to release planning. Hybrid approaches are focused on combining well-structured widespread methods with agile principles [7]. However, reality shows that the synchronization of these approaches does not work successfully [8]. All of these approaches have their own specified release plans that best fits their project characteristics. Yet, there is no established means in science nor industry resolving these differences on handling release planning in an automotive context that have to struggle with different development practices. Therefore, there is a lack of synchronization for a coordinated release plan of all involved stakeholders. Although there are release planning approaches that mention stakeholders as a fundamental impact factor that has to be considered [9] there is no further recommendation and sufficiently detailed consideration. The literature study by Ameller et al. [10] indicates that the 17 selected papers analyzed therein do not go into detail about stakeholders as an essential factor of input. The stakeholders are an important part of planning and have a major influence on its creation and usage [11]. The study of Marner et al [12] states it is not obvious which stakeholders are to be affected and what benefit they expect from a release plan. To support release planning, it is necessary to create transparency regarding the parties involved and the required content. The stakeholders of a release plan, the kind of content as well as the served purpose have to be identified and are an important aspect of a release plan. For this reason, we focus on stakeholders and consider in detail not only who the stakeholders are, but also what content they need from the release plan and what purpose they serve. These three points form the research question this paper is based on. The results presented here have been produced in cooperation with two pilot projects of one German automotive original equipment manufacturer (OEM) and are based on the scientific methodology of action research. Furthermore, this paper contributes to the discussion, which stakeholders have to be observed to set up a target-oriented and structured release planning. Together with existing release planning methods, our results can improve multidisciplinary release planning considerably. We present first empirical evidence of two pilot projects performed in industry.

The remainder of this paper is structured as follows: Related work is presented in Section 2. Section 3 defines the research approach including the research questions and design, the research site and the participants, the data collection and analysis procedure, as well as the data collection instrument. The results are reported in Section 4. Similarities and differences of the results as well as a discussion are part of Section 5. Section 5 includes learnings and recommendations for other companies. We conclude our work and outline future research in Section 6.

## 2  Background and Related Work

**Hybrid project environment: soft-and hardware development with corresponding development methods**

In the automotive domain, vehicles are developed in a hybrid project environment that is given with two conflictive parts. On the one hand, there is the strategic framework with processes consisting of an idealized and generic description for a product development process [13]. This framework defines different phases of the development process and characterizes the strongly regulated automotive industry. It contains several milestones planned a long time in advance before vehicle projects related to production and distribution start. All the defined and fixed milestones given by the company can be summarized as a time horizon and must be followed. The strategic framework has to be considered at the beginning of the release planning process and restricts the possible levels of flexibility in the development process. On the other hand, there is the operational level where projects are developed with different development methods depending on each project's characteristics. On this operational level, the projects can be developed in an agile, hybrid or traditional way. The results of the large-scale online survey show the current state of practice and prove that agile, hybrid and traditional development methods are used in industry [14]. There are different possibilities to do planning but there is no general approach for structured release planning that brings together different development processes effectively and that serves the strategic framework. Due to the fact that soft- and hardware development is a complex and challenging task involving many stakeholders [15], there have to be different ways to do planning because of the projects' characteristics. In the automotive industry, there are different development levels, which are characterized by function development at the lowest level over ECU integration up to the vehicle level. Depending on the individual mandatory safety and security level of each software, different characteristics of release plans are required additionally. As a result of the complex specifications and influencing factors, flexibility with regard to release planning and consideration of certain stakeholders is decreasing. Nevertheless, all the different ways have to work together well in one vehicle and therefore a general structure for a coordinated release planning is necessary.

New ways to address projects and development processes are possible due to industrialization 4.0, which offers new possibilities for networking and collaboration [16]. To use these new technologies and to integrate the vehicle seamlessly into the digital urban world, a new type of vehicle electronics network is required. An end-to-end electronic architecture forms the technological basis for implementing future innovations [17]. To consistently apply this system architecture new high-performance computing platforms are necessary. The goals of future electric/electronic architectures are open and scalable architectures that are sustainable and durable. They are expected to be able to map current megatrends such as autonomous driving and connected cars, as well as future challenges such as swarm intelligence and artificial intelligence [18]. For this reason, ECUs and functions that are currently distributed throughout the vehicle are grouped on one or more computers. This centralizes and maps the complexity and responsibility of several ECUs in one high-performance computing platform (HPCP). A HPCP thus hosts hundreds of different functions, e.g. from the areas of connectivity, driver assistance, energy and charging, which makes the requirements, planning and control of that ECU much more complex than current vehicle architectures. We present the identified stakeholders of release plans for such a new HPCP and for one function located on it.

**Release Planning Approaches**

The consideration of release planning combining software and systems engineering is hardly to be found in the literature. Sax et al. [19] conducted a survey on the state and future of automotive software release and configuration management. The outcome of that study faced challenges during release development and management. The focus of this study was set up to technical trends as variant management, software updates and multi-domain system design. In their findings Sax et al. cover both software and hardware aspects. Nevertheless, their results are strongly aimed only at future technical challenges. Müller et al. [20] elaborate fundamental requirements for the IT support with regard to the release management process. They presented to which extent product data and process management technology meet their identified requirements. Based on their gained experience they formed four categories (IT infrastructure, process control, process enactment support and usability) for IT support. The requirements shown by Müller et al. refer to the automotive context and they uses release management as an example in their work to gain requirements for an IT tool.

In addition, release planning in agile software development projects for both single projects and for scaled projects is discussed in the literature. Heikkilä et al. [21] presented research of release planning in large-scale agile software development organizations. They conducted a longitudinal study and identified a gap of firm empirical evidence of results demonstrating practices for scaling up Scrum with regard to release planning. They filled this gap by presenting two results produced within a case company using the Release Iteration Planning method [22]. However, their results focus only on software development in an agile context and did not include the hardware development. Danesh et al. [23] performed a multiple case study to investigate the methods companies use to plan for new software releases. Their results demonstrate that experienced companies prefer improving their existing software products rather than creating a new plan. These results correspond partially with the challenges identified by Marner et al. [12], who has highlighted the problems that exist in both software and hardware development regarding release planning within an OEM.

Heikkilä et al. [24] describe in their case study how release planning is done in a large-scale Scrum development organization. One of the challenges identified in this study were the over commitment caused by external pressure. They even presented benefits such as increased flexibility and decreased development lead-time. Heikkilä et al. [25] used a case study to improve coordination of work of multiple agile development teams. They tested two joint release planning events to set up the case companies' own joint release planning events.

To the best of our knowledge, there is no explicit research on release planning in co-existing traditional and agile way of working in the automotive domain or in a related domain with similar boundary conditions. Few publications deal with agile release planning, but none of them regard the targeted hybrid project environment combining software and hardware engineering. Karvonen et al. [26] analyze both direct and indirect impacts of agile release planning practices within their systematic literature study. In addition, Ameller et al. [10] conducted a literature study to report on software release planning models. Both publications focused only on software engineering.

The results of a systematic review on strategic release planning models [10] show that there are various kinds of mathematical models and simulations that are unusable in practice [27]. By using these approaches in daily work, the practitioners reported that they are either too simple to generate a benefit or so difficult to use that they cannot reconstruct the structure and procedure [28, 29].

**Release planning approaches addressing influencing factors**

The approach EVOLVE [30] and its extensions [31] are a support for decision-making of software release planning. This method contains the strength of genetic algorithms and the flexibility of an iterative solution method. EVOLVE in its original approach takes into account numerous constraints such as software requirements, stakeholders' priorities, prioritization of requirements by stakeholders and effort estimation for all requirements. The extended version EVOLVE II incorporates additionally soft requirements and objectives into their proposed decision-making process. EVOLVE II concentrates with its approach on software release planning taking into account numerous influencing factors but excludes the hardware development.

Saliu and Ruhe [11] describe ten key technical and nontechnical aspects impacting release planning. They refer to the aspects as dimensions. The following selected dimensions particularly affect the automotive industry because these dimensions are what compose the complexity: time horizon, objectives, stakeholder involvement, prioritization mechanism and technological as well as resource constraints. Lindgren et al. [32] used the key aspects of release planning of [11] and performed a multiple case study in the context of software and system development projects. Furthermore, they considered the state of the practice for release planning in industry. Within their performed work, they validated the defined aspect of [11] involving industrial companies and proposed one further key aspect (short- and long-term planning). They stress that too little attention is dedicated to the complex topic.

Both [11] and [32] assume the availability of or knowledge about certain constraints such as functional requirements or stakeholder preferences. Within their work, there is no closer proposal to look at or how to identify which constraints are necessary or important. However, in such a complex environment as the automotive domain with many technical dependencies and fixed milestones, it is not always obvious what kind of constraints are given and what has to be considered. Based on our own experience we give some additional examples of different constraints especially for the automotive industry below:

- prototype cars and target hardware (on a certain development level the prototype car environment is needed for final testing and calibration)
- climate conditions for testing (many functions depend on climate conditions; that means the seasons have to be taken in account to avoid expensive costs)
- complete vehicle maturity level (the required maturity level of all functions has to be fulfilled on a certain degree)
- safety and security aspects (functions that have to fulfil lots of safety and security aspects need much more time for development and testing in advance)

Extracting out of the analysis results of [10] within their literature study, it can be stated that there are incomplete input factors processed by the analyzed models within their study. For release planning in the automotive context, further factors (safety and security, practical applicability, and requirements given by the company), has to be included as input in the models. None of the analyzed approaches deal with input factors such as safety and security requirements or a more complex list of fixed requirements given by the company to fit industry needs. Safety and security aspects play a major role within the automotive domain [33]. Another important finding of the analyzed publications of [10] is the lack of practical relevance. Only one method developed by Heikkilä [25] is related to industry. There is no further validation performed in industrial settings as well as further literature except [11], [32] dealing with aspects a release plan should include. All other evaluated approaches have been generated with academic references.

With this paper, we rely on the input factor *stakeholder involvement* in accordance to [11] and take them as a reference to detail this factor extending them to the automotive context. This key factor is crucial and has a huge influence on the content of a release plan in the automotive industry. For this reason, we focus on relevant stakeholders of release planning and show which information stakeholders provide in the release plan and which information they require from a release plan.

## 3 Research Approach

### 3.1 Research Questions

Stakeholders of all kinds are very important for creating realistic and goal-oriented release plans. Their requirements have a great influence on the development of a release plan and have to be aligned with the defined guidelines of the companies. Therefore, it is important to understand who the stakeholders are and what they require from a release plan.

This paper aims to answer the following research question (RQ 1-3) in **Table1**.

*Table1. Overview of the research questions*

| | |
|---|---|
| RQ1 | Who are the involved stakeholders? |
| RQ2 | What kind of content do the stakeholders require from a release plan? |
| RQ3 | What benefits do stakeholders gain from the information in a release plan? |

**RQ1:** This question aims to identify the involved stakeholders of a release plan. Specifically, we study the relevant stakeholders of two pilot projects and therefore of two perspectives release plans are made for.

**RQ2:** After identifying the relevant stakeholders of a release plan, it is necessary to know what kind of content the stakeholders require from a release plan or which information the stakeholder can provide for the plan. That is what we investigate with this question.

**RQ3:** This question aims to determine what purpose stakeholders are pursuing and why they require information contained in the plan or why they providing information in the release plan.

### 3.2 Research Site and Research Projects

The results were developed at Dr. Ing. h. c. F. Porsche AG, a manufacturer that builds sports cars for everyday driving. The division EE (Electrical/Electronic) and one further division within Dr. Ing. h. c. F. Porsche AG in Weissach, Germany are responsible for the development process of electronic systems and its integration into the development process of the complete vehicle. For achieving this goal, transparent development processes and hence accurate release planning are essential.

The findings were generated for a HPCP and one function located on it as pilot projects within the case company. The first pilot project is one of the new system architectures as mentioned in section 1 and follows a traditional development procedure. That HPCP can be regarded as a representative example, since it provides a main ECU within the network. The second project is a function located on this platform and is called predictive thermal management function. The function is one of several software functions that are located on a HPCP and is developed in an agile way. The goals of this function are to increase performance, efficiency and comfort through predictive thermal management by cooling and heating of the battery. The two pilot projects represent the hybrid project environment consisting of traditional and agile development methods. Furthermore, the hardware level is included in the results by the HCPC. There were no existing release plans for either projects, as they represent new technologies. In addition, the selected HPCP is a joint group development within the Volkswagen Group and this fact increases the complexity. Due to its structure, the selected function is also a highly distributed and networked function. For these reasons, the results listed in this work are appropriate.

### 3.3 Research Design

To answer the research questions, we selected a two-step research approach and conducted action research proposed by Myers [34] in a series of workshops (Fig.1). Action research is selected because of a prevailing real particular problem situation to develop new release plans for two industry related projects. The objective of step one is the definition of relevant stakeholders (RQ1), the identification of expected content for the stakeholders (RQ2) as well as the gained benefit for the stakeholders (RQ3). The evaluation of the results for the two projects were done in step 2.

**Step 1: Data collection and analysis procedure**

The identified challenges in release planning of the case study of Marner et al. [12] served as input for step 1. The results for each RQs is carried out in a cyclical process as demonstrated in Fig.1. With feedback loops, the researchers planned changes to be introduced in the case projects, took out these changes, reflected on them, and evaluated the results. We repeated this cycle several times for each research question. Step 1 lasted from April 2019 until July 2019.

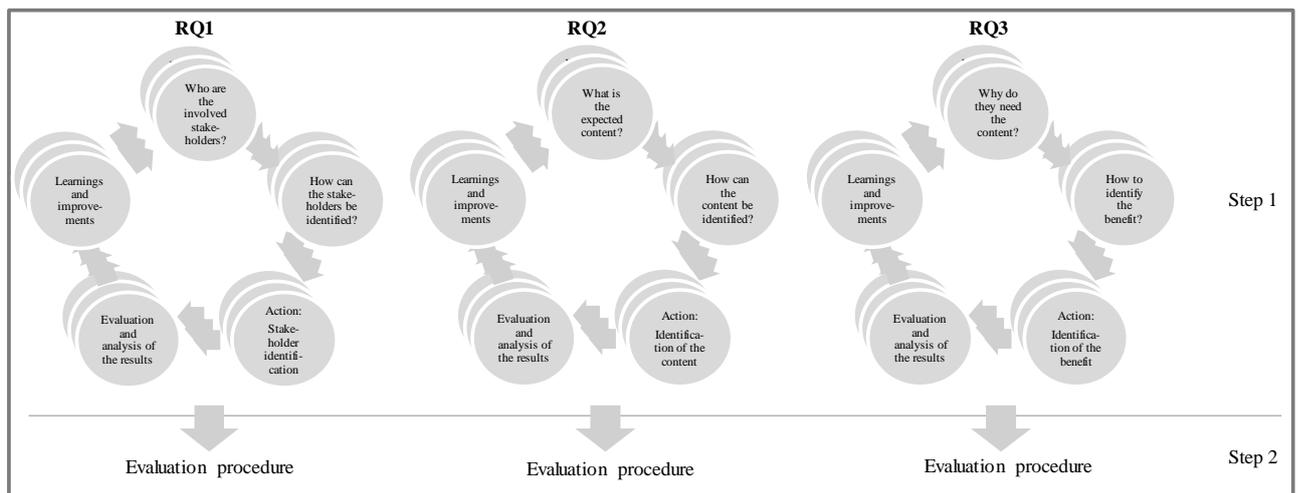

*Fig. 1. Used procedure for both research projects*

The action research team consisted of researchers and participants who collected the results. The researchers were the same persons for both projects, only the participants differed. The researchers are the three authors. The participants of the action research team of the HPCP consist of one HPCP owner, three developers, one tester, one project owner and one representative of the quality department of the HPCP. The action research team of the second project is composed of one software function owner, three developers, two testers and one responsible from the quality department. As reference group, two department heads and one integration manager supported us for both projects. Among the participants of the action research team as well as the reference group were stakeholders who are identified as stakeholders in the results. We combined the experience of the authors due to the active involvement of the first author in the automotive software development and the experience of the second and third author regarding software engineering.

The action research team for step one was composed of participants all working at Dr. Ing. h. c. F. Porsche AG. The results were carried out in close collaboration between researchers as well as practitioners and both benefit from the findings. The results for the HPCP were worked independently of that of the predictive thermal management function

The action research cycles, started by diagnosing well known still unsolved problems of relevant stakeholders emerging from [12]. After a procedure to answer the problem was found, this was worked out in the respective pilot projects.

The involved stakeholders (RQ1), the required content (RQ2) and pursued purpose (RQ3) with regard to the release planning was specified in that way. The results generated were revised after each cycle by the first author. Emerging issues, such as vague phrases, were addressed before further cycles were conducted. The results from each cycle were incorporated into later ones. We used the post-processed results as new inputs for incremental improvement systematically. The results were created in an iterative way to confirm its design and to draw a more complete picture of the way release planning was done by different participants. This allows for more validated results, but gave every participant the chance to provide further qualitative results by sharing their experiences.

**Step 2: evaluation procedure**

In Step two, the results applied in both case projects were analyzed and evaluated with further participants between July and September 2019 with experts from Audi AG. After achieving the results for the HPCP, four workshops over 60 minutes were conducted. Also to these workshops, we invited the experts due to their knowledge and experience gaining criticism. These workshops were held with project managers of ECUs and further HPCP owners working at Audi AG to align the results of both brands on this complex topic. In these workshops, the generated results were presented and discussed. The provided feedback is incorporated into the results presented in this work.

The validation of the results for the pilot function was done in form of two 60 minutes meetings with attendance of the corresponding partners from Audi AG, who are responsible for function development on their side. The received feedback was incorporated into the results in order to achieve improvements by using the knowledge from Audi AG.

## 3.4 Threats to Validity

We used the following four criteria suggested by Wohlin et al. [35] and we refer specifically to action research in accordance with Staron [36].

**Construct validity.** This validity is related to the design of our study and concerns to the diagnosing phases as well as the planned activities. Based on the study by Marner et al. [12], indicating that stakeholders of a release plan are not immediately apparent, we have ensured a common understanding of the problem or research question at the beginning of each research cycle. In the process, we have incorporated interpretations from each individual and created a common understanding. During the planning of actions, we avoided relying only on one's individual opinion about the improvements. To report the results we did it together in common workshops.

**Internal validity:** Internal validity with regard to action research focus on action taking. Because of the close cooperation, which was spread over several cycles and lasted for a certain period, the industrial partners are learning. As a result, their assessments are no longer as objective as they were at the beginning of the project. We reduce this threat by performing reviews with participants from different departments within the case company and from Audi AG.

**External validity:** Due to the fact that action research takes place in a specific context, we have to reflect on the impact of our results. As the results only represent one specific case, it might not be possible to generalize them. However, the fact that the case company has the same framework conditions (regulated domains, complex supplier relationships and high safety requirements) as similar OEMs, others could benefit from the developed approach. Furthermore, the selected projects represent an ECU still in development and a software function not yet published.

**Conclusion validity:** Conclusion validity is reflected amongst others in the evaluation phase. There is bias concerning drawing conclusion because of the fact the action team is part of taking the action. If you have to judge your own actions, you are simply less objective than judging results of others. Once we realized that our results revealed no differences, but that we still recognized effects on our actions, we took this as new input for the next cycle. We also included other participants when the action team could not find a common consensus. Here, people in the role of technical department heads were particularly suitable because of their specialist knowledge and their ability to look at things from a distance.

## 4 Results

In this section we present the results following the research questions (Section 3.1). The results will be reported for each of the two research projects. We start with the presentation of the results of the HPCP and afterwards the results of the software function will follow. For a better understanding of the results below, the definition of terms has been made as follows.

**strategic framework:**
specification of the OEM representing the time and content requirements; consisting of project-specific milestones, vehicle specific milestones and release dates

**implemented content:**
description of the content to be developed, what will be implemented in detail at what time (e.g. at different levels possible)

**delivery date:**
time at which a certain scope must be delivered in (required maturity level) containing the agreed implemented content

**required maturity level:**
specified maturity level by the OEM, which is integrated in the product development process and which every hardware or software project has to observe

**planned maturity level:**
short description, which scope will be developed related to a maturity level (wording that is management like)

These definitions will be used in section 4 but mainly applied in section 5.

## 4.1 Involved Stakeholders

In the following, the identified stakeholders will be presented and explained in detail along with their role description.
*RQ 1. Who are the involved stakeholders of a release plan?*
First, the results of the HPCP are presented. The results of the software function follow afterwards. The identified stakeholders of the HPCP are classified into three categories: management, development and after sales and market. (**Table2**)

*Table2. Overview of the identified stakeholders of the HPCP*

| category | management | development | after sales & market |
|---|---|---|---|
| **stakeholder** | ▪ project manager | ▪ function owner | ▪ smart mobility department |
| | ▪ model series organization | ▪ functional safety department | ▪ after sales representative |
| | ▪ integration management department | ▪ type approval department | ▪ HPCP quality representative |
| | ▪ cross section department | ▪ test manager | ▪ production & logistic department representative |
| | | ▪ software supplier (basis software) | |
| | | ▪ hardware supplier | |

### 4.1.1 High-performance computing platform

**Management**
Stakeholders having a management role, bearing responsibility and having decision-making authority, define the category *management*.

The first identified stakeholder is the *project manager* responsible for one HPCP. The project manager is in charge from the definition phase to the handover of responsibility to production. In addition, this role monitors the component development according to the specifications, the process-safe manufacturing process as well as the planning, tracking and regular reporting of these components. This role description also applies to suppliers of hardware, software or third parties who perform the same tasks for their scope.

The companies in which these project managers are employed and especially OEMs have a line organization. The authority to give directives carries out from top to bottom. This also enabled department heads and even their superiors to be identified as stakeholders in the *management* category.

An organizational special feature at the case company is the so-called *'Model Series Organization'* [37]. A *model series* has the corporate responsibility for its vehicle scopes as well as platform scopes of all co-users. It is therefore responsible for the cross-departmental compliance with all specifications regarding technology, deadlines, costs and quality. It includes the cost targets for purchased parts (direct material costs and investments), the development costs and the target achievement of the production parameters. The required maturity levels are defined by the model series. For this reason, the model series organization is an identified stakeholder of the release plan and is interested in specific content of that plan.

From a planning and decision making point of view, *integration management department* plays an important role in the vehicle release process. It ensures the integration of functions into a vehicle by tracking and evaluating the corresponding maturity level. It contains as well the interface between development and model series. Integration management is also involved in the definition of maturity levels.

The next identified stakeholder is a technical connection between project manager and integration management. The *cross section department* specifies general requirements for ECU development. For example, this contains requirements for the used bus technology like Ethernet or requirements for diagnosis.

**Development**

The second category, *development*, represents stakeholders who are involved technically with the HPCP. This category includes the development and testing level.

In the HPCP, there are numerous functions. Such functions are the *function owners'* responsibilities. A software function is the smallest unit delivered to a HPCP. Such a software function includes the entire effect chain from the sensors and actuators, to multiple HPCP domains, across the entire network. The effect chain ranges from electronics to display or backend functions.

Each software function is integrated into the overall software of a HPCP. For each software function there is a responsible person. Therefore, the function owners are involved stakeholders of the release plan.

*Functional safety department* and *type approval department* have great impacts on a HPCP. Functional safety [33] refer to risk reduction due to malfunctions of electronic/mechatronic systems. Type approval deals with obtaining the required operating permits for the market launch and registration of customer vehicles in sales markets. These two areas influence the creation and implementation of the plan regarding fixed milestones and testing. For this reason, these departments have to be considered as stakeholders.

Vehicle testing is assigned to this category. It serves to ensure durability, functionality, reliability and quality throughout the entire product development process as well as during the first months of the current series. The fulfillment of customer- and market-specific requirements is checked with different standardized or demand-adapted test methods. The *test manager* as an identified stakeholder is responsible for the preparation and execution of the test activities.

The *hardware supplier* is another important stakeholder and develops the necessary hardware for a HPCP. This supplier is responsible for hardware development, integration of (all) software, product validation (testing) and production of the computing platform.

The *software supplier* is responsible for development and delivery of the basis software of the HPCP. This stakeholder ensures the basis functionality of HPCP and provides attributes such as diagnostic capability and networking. Both the hardware and the software supplier work closely with the project manager to ensure the most efficient development.

**After sales and market**

The third category, *after sales and market*, includes stakeholders that are not permanent involved to research and development, but contains key aspects in the vehicle's value chain.

The *smart mobility department* is concerned with the alignment and operationalization of the product and portfolio strategy for connected car services, including lifecycle management. This department as an identified stakeholder handles the creation and planning of connect services.

The *area after sales* is responsible for servicing the customer vehicles in the worldwide OEM workshops and thus represents the most important point of contact after purchasing the vehicle.

The *quality department* is responsible for controlling all quality issues within the model series for the development process and series support. This department with *HPCP quality representatives* for each model series ensures the requirements in the development process of electronic/electrical components, as well as defining and implementing the quality objectives in the vehicle specifications.

The department *production and logistics* for a special model series includes the series production of all vehicles. Even in this department, there are representatives who ensure that their model series meet the requirements with regard to production and logistics. The high quality and punctual completion of the vehicles across all trades is closely connected with the planning and control of the production and logistics processes.

Following the action research cycles, it was revealed that purchasing is not a stakeholder in the plan of a HPCP. This is related to the fact that purchasing is active before the development phase.

### 4.1.2 Function – predictive thermal management

The identified stakeholders involved in the release planning process of the function are grouped into three categories (**Table3**).

*Table3. Overview of the identified stakeholders of the predictive thermal management function*

| category | Agile Team | technical stakeholders (functional interfaces) | informative stakeholders (no functional interfaces) |
|---|---|---|---|
| **stakeholder** | ▪ development team | ▪ provider predictive data | ▪ management |
| | ▪ product owner team | ▪ base thermal management | ▪ production department representative |
| | | ▪ high performance computing platform | ▪ software quality representative |
| | | ▪ components of coolant circuit | ▪ integration management |
| | | ▪ software supplier | |

**Agile Team**
The *agile team* itself forms the first category. The agile team is composed of the *product owner team* and the *development team*.
The members of the agile team and their corresponding role descriptions are also interpreted and performed at Dr. Ing. h.c. F. Porsche AG as described in [38]. For this reason, we will not discuss these role descriptions in more detail and will not look into them.

**Technical stakeholders (functional interfaces)**
The second category called *technical stakeholders* consists of stakeholders having functional interfaces and therefore an impact on the release plan. This includes functions that either deliver input data or receive data from the predictive thermal management. The required and planned maturity level of all these functions are dependent on each other and raises the common coordination effort. In addition, the technical stakeholders having direct impact on the software itself are located in this group. The first role within this category are the *providers of predictive data* for the predictive thermal management. There are various basis services that process raw data from different sources, combine these predictions and make it available as preprocessed prediction data for different user functions, where the predictive thermal management is one of it. Each of these basis services aims at the prediction of special use cases. This is why many different of these services have to be taken in account. Depending on whether the service is located on the same as the pilot function or on different HPCPs or even on the backend, this has a major impact on the release planning activities and raises the complexity.

The *base thermal management function* is another stakeholder that acts as a backup function for the predictive thermal management when there are no predictive data available. The base thermal management function provides sensor data, controls the actuators and serves as a monitoring and diagnostic device for the actuators and sensors of the coolant and refrigerant circuits. The base thermal management transforms the output of the predictive thermal management in order to control the actuators. Furthermore, the maturity level of the predictive thermal management is highly dependent on the base thermal management. For this reason, a regular and tight coordination is essential.

The *HPCP* is the corresponding hardware platform the predictive thermal management is located. The HPCP is responsible for the whole integration process between software and hardware level. That includes the build process of all software functions located on this platform. Especially functional changes that include new interfaces must be incorporated right from the start; otherwise, the lead times will be too long.

The *components of coolant circuit* include all devices that need to be cooled or heated by the thermal management system. These devices have a great need for the maturity level planning to be coordinated and known at an early stage, as the proper use of the devices depends largely on the thermal management system. Conversely, the sensor data of the devices are a prerequisite for the predictive thermal management. Due to this interdependence, it is necessary to determine very precisely when the corresponding maturity levels are reached for both sides.

The *software supplier* defines the build process and due to that specifies the period, how much lead time has to be planned until delivery to the HPCP. Furthermore, when drawing up the release plan, the available resources at the supplier must be kept in mind. The software supplier receives the required maturity levels from the OEM and has to adopt his build process to these given milestones.

**Informative stakeholders (no functional interfaces)**
A group that has no functional interfaces but is of a purely informative nature characterizes the last category of the identified stakeholders for the pilot function.

The first stakeholder within this category is *the management*. Amongst others, their tasks include decision-making authority and budget responsibility. Furthermore, it is the responsibility of the management to maintain the software conformity and for this reason; they are interested in a comparison of target and actual maturity levels. The main focus is the monitoring whether the function remains on schedule. However, they are also the stakeholders of all kinds of conflicts that have not yet been solved and have the task of resolving the conflict at their level.

*The Production department* contributes to release planning with its own process**s**, in which the sourcing process should be incorporated, as milestones arise retroactively. *The production department representative* is interested in ensuring that the functions relevant to production have the necessary degree of maturity.

Following the workshops, it was revealed that purchasing is not a stakeholder in the plan of a software function. This is related to the fact that purchasing is active before the development phase.

The *software quality department* specifies quality related requirements that have to be considered and addressed in planning. In addition, this group monitors specific quality related milestones. Another task is the execution of software supplier audits. The results of such audits have to be integrated into the release plan as well.

The *integration management* is the last stakeholder within that category and is responsible for transferring the required milestones defined by the OEM to a completely specific vehicle project. This stakeholder specifies the required maturity levels and conducts a comparison of target and actual status of the developed software.

### 4.2 Content Needed

This section deals with the relevant content for the identified stakeholders. RQ1.2. *What kind of content do the stakeholders require from a release plan?* First, the results of the HPCP are illustrated based on the structure of the defined three categories. The required content of the stakeholders for the software function is outlined afterwards.

#### 4.2.1 High-performance computing platform

**Management**
The *project manager* is the main stakeholder of this plan because this role requires most information or content. Its task is to consider the development comprehensively and therefore s/he is the contact person for all further stakeholders. The automotive domain is a strongly regulated domain. The strategic framework with fixed milestones and required maturity levels are given by the development process of a vehicle. The project manager takes this strategic framework as a basis and includes further contents such as delivery dates from suppliers of hardware and of software. Furthermore, the project manager adds points in time to this plan with regard to delivery dates. Specific deadlines for the basis software of the HPCP has to be included into the plan as well. The base software is the software on an ECU that provides the base functions. Software functions located on a HPCP are not part of the basis software. Different software version deliveries and required maturity levels of the software functions and of the basis software are others elements of the release plan. Additional information such as functional safety dates as well as delivery dates are part of the plan. Functional safety dates represent additional dates that has to be observed additionally if the product to be developed has to operate the Functional Safety [33].

In general, it can be stated that the information all other identified stakeholders need is a subset of the information the project manager creates and requires.
The amount of information required from the software and hardware suppliers is similar to that of the project manager. These stakeholders need the delivery dates for hardware and basis software as well as the required maturity level for hardware and basis software.

Information for *department heads* and supervisors is significantly less and more compact. On the one hand, they need reports with traffic light status and on the other hand the amount of software functions that have to be integrated for every software release of the HPCP will be indicated.

The *model series organization* demand for information in form of a report with traffic light status and the planned maturity levels of hardware and software. They also need the implemented content for hardware and software.

The *integration management* need a set of information containing the delivery dates of hardware and software as well as the planned maturity level regarding to software and hardware. The type approval dates are required information, too.

The *cross-section department* needs information about the implemented software scope and the required maturity levels of the HPCP. In this context, software scope contains the basis software and all integrated software functions. In addition, information regarding hardware and software delivery dates are needed by the cross-section department.

**Development**
*Function owners* need the following information of the plan consisting of strategic framework and delivery dates with regard to hardware and software. Deadlines for the basis software of the HPCP are additional information this role requires. Furthermore, the function owners need required maturity levels of its specific software functions and the required maturity level of the corresponding hardware. The function owners provide the implemented content of their software.

The next group within this category is formed by type approval and functional safety representatives. The area *type approval* is interested in specific technical information as dates of type approval. This area is responsible for all kind of technical compliance, communication with authorities and technical guidelines. Contents that are subject to *functional safety* require delivery dates for hardware and software. That strongly relates to the ISO 26262 [33]. Type approval and functional safety both need the same information from the plan as the planned maturity level of hardware and software as well as the implemented content of hardware and software.

Development of software and hardware may have severe effects on interfaces, which is why tests have to be done. The *test manager* handles the testing activities with the delivery dates regarding hardware and software. Moreover, this department needs the implemented contents for software as well the type approval dates.

For the development of the ECU, the *hardware supplier* first needs information from the strategic framework such as project-specific milestones and release dates. The hardware supplier also requires additional required maturity levels of hardware specified by the OEM from the release plan for the further development of the HPCP. In coordination with the hardware supplier, the project manager also specifies the delivery dates for hardware.

The *software supplier* is responsible for developing the basis software, containing the operating system, hardware near drivers and system functions such as standard diagnostics. The supplier of the basis software also requires general information such as project-specific milestones and release dates for its development. Similar to the hardware supplier, the software supplier aligns the development of the basis software with the OEM's specified required maturity level. In addition, this supplier is further in charge of the integration of software functions for function demonstration. In collaboration with the project manager, the delivery dates for the basis software are also agreed and included in the release plan.

**After sales and market**
The responsible persons of the *smart mobility department* need the strategic framework and specific project milestones. In addition, they demand for the diverse planned maturity levels of hardware and software.
The *after sales* department is more interested in delivery dates for hardware and software after start of production. This department needs even a mapping of software and hardware compatibilities regarding the HPCP.

From a *HPCP quality management* point of view, the project milestones of the release plan are necessary. Furthermore, they require the planned maturity level of hardware.

The specific delivery dates for hardware and software of the HPCP and even the project milestones are needed for further processing by the *representatives of production and logistic*.

### 4.2.2 Function – predictive thermal management

**Agile Team**
The function owner is the first representative of the *agile team* and this role is part of the *development team*. Within the project, the function owner is assigned the task of the central contact person. This person gathers the necessary information for the software project given by the company, comprising the following points: the strategic framework forms the time horizon and contains several milestones e.g. strategic project milestones, vehicle specific milestones and release dates, given by the company. The task of the function owner includes the transfer of these milestones to the corresponding specific project.

Furthermore, its tasks consist of the transfer of the delivery dates regarding the HPCP as well as the transfer of the agreed delivery dates of the software supplier into the release plan. For this reason, the release plan is an important base consisting of predefined dates and corresponding content for development as well as the application done by the software supplier. All milestones related to test activities are a significant part of the plan and must therefore be reflected in the plan, which belongs also to its task. Further additional dates result from the detailed implemented content of the software and sub functions concerning the upcoming release and has to be done by the function owner. In that case, the function itself is divided into several sub functions providing a more detailed level of description.

The required maturity levels given by the strategic framework for the function and its sub functions have to be transformed in order to plan the individual content of each release. The content of each release needs to be planned in the way that the demanded required maturity levels can be fulfilled.

**Technical stakeholders (functional interfaces)**

Within this category, three members consisting of *provider of predictive data, base thermal management* and *components of coolant circuit* all need the same information out of the plan due to their technical affiliation. This information includes, for example, requirements for the required maturity level of the common technical interface and vehicle specific milestones for different complete vehicle testing dates as well as testing locations derived from the information provided in the plan. Additionally, these technical stakeholders also demand for delivery dates for its sub functions.

There is a close exchange between the software function and the *HPCP* ensuring the integration of the software on the hardware without difficulty. Therefore, the hardware demands the delivery dates from the software. Furthermore, the HPCP requires the implemented content of the software due to given milestones. In addition, the exchange includes an agreement on technical data such as the required resource reserve (RAM, ROM, CPU etc.) of the software that have to be considered in the release plan.

The *software supplier* necessitates boundary conditions for the implementation of the scopes to be developed. These conditions include as well the given milestones by the OEM and time dates when the software has to be released defined by the delivery dates of the HPCP. The implemented content of each release needs to be clarified together with the software supplier in order to respect the development capability of the software supplier.

**Informative stakeholders (no functional interfaces)**

The *management* requires information contained in the plan due to top-level management summary that is prepared for management purposes. This stakeholder is interested in a status report of the project and needs no detailed or specific content. The next stakeholder within this category the *production department representative* needs general project information such as planned maturity levels and the implemented content for the sub functions and the whole project runtime.

The *software quality department* specifies specific quality-related maturity levels and expects providing planned quality-related maturity levels. This stakeholder is not interested in general maturity levels but focuses maturity levels due to software quality.

The *integration management* provides required maturity levels of software and demands planned maturity levels. If any deviations between planned and required maturity levels occur, this stakeholder can cause an escalation in order not to endanger the development process.

**4.3 Benefit and Purpose of the Information**

In the following it is presented why the identified stakeholders need the information from the corresponding release plan. *RQ1.3. What benefits do stakeholders gain from the information in a release plan?*

In summary, it can be stated that all the contained information of both release plans can be used as an input for the involved stakeholders. The input extracted differs in the way it is used by stakeholders. On the one hand, the input is purely informative and on the other hand, it provides basis information for the further development process. The benefit and purpose of the implied content can be summarized because the stakeholders gain similar additional value. The results of the HPCP are discussed first, followed by the findings of the software function.

**4.3.1 High-performance computing platform**

Three stakeholders in the category *development* (*function owner, hardware supplier and software supplier*) are key stakeholders who both require some information from the plan and provide information to the plan. This group uses the information for its respective development, whether it is hardware, software and basis software. Without these stakeholders, the development of the HPCP cannot be applied and implemented in a target-oriented way.

The information contained in the release plan is used for the four stakeholders of the category *management*: *project manager, model series organization, integration management* and *cross section department*. This group needs the defined information for project tracking, costs monitoring, content overview and different deadlines. Every stakeholder that requires a status report in form of traffic lights deploys the report for tracking and an escalation instrument regarding the HPCP.

In close exchange with the *HPCP* the *development* and the *cross section department* uses the information for planning test activities (e.g. for network tests and tests for interfaces) and for target comparison regarding to production relevant content.

The *after sales department* uses its information to establish an after sales strategy. The information requested will also be reused for the organization and planning of the spare parts warehouse.

Several stakeholders need the different type approval dates. Function owners adjust their planning to these dates. With the given information, they are informed when they are no longer allowed to change anything in the development. *The integration management* takes on a superordinate role with regard to these deadlines. The integration managers look at the technical departments and sensitize them to the dates. On the one hand, the department for *type approvals* and *functional safety* coordinates with the development department in order to announce the dates. On the other hand, they obtain information from the development department when they intend to implement their functions.

The required maturity levels are key input for overall vehicle planning. These levels are indicative to develop high quality software as well as hardware and therefore important for almost all stakeholders involved in this release plan.

In the area of *production*, the contents of the release plan are used to fulfil the project specifications related to plan logistics. Furthermore, the information are needed for the synchronization of the testing equipment.

### 4.3.2 Function – predictive thermal management

Similar to the HPCP, certain stakeholder groups pursue common purposes with the information provided within the release plan.

The *technical stakeholders* have to generate their own planning and for that, they need delivery dates, required and planned maturity levels and implemented content of the sub functions from the release plan. Using this information given in the plan, they prepare content to be implemented in its function and its sub functions. The *technical stakeholders* and the function owner coordinate their individual plans to achieve the agreed targets of different delivery dates as well as the maturity levels. These stakeholders also create a detailed plan of the implemented content of sub functions to synchronize technical interfaces.

The *agile team* including the function owner is both responsible for the plan itself and for the development of the implemented content as well as the required and planned maturity levels. For this reason, they need all the information provided by the plan. All the information contained in the plan comes together in the agile team to monitor, plan, prioritize and report on the entire software development process.

The category *informative stakeholder* consisting of *management*, *production*, *software quality representatives* and *integration management* need different information from the plan, but they pursue the same purpose as the *HPCP* with the information received. The *informative stakeholders* only demand information that is not processed in another plans, but to monitor the progress of the corresponding software development project. They have no responsibility for their own software, but rather have coordinating and monitoring tasks to perform. These stakeholders do not only pursue particular software but also have an overall focus on several software functions and consider the project to be developed holistically.

### 5    Comparison and Discussion of the Results

Having presented our results, we now will discuss the results.

Stakeholders are an essential influencing factor for release planning, as they provide information as an input factor for the planning and require information from the developed plan. Stakeholders strongly depend on the use case as the results in Section 4 have already shown. Although both pilot projects come from the same project context, different stakeholders are needed for release planning of the HPCP as well as the software function at working level for software development. Nevertheless, there are differences besides some common categories of stakeholders. Due to content, even there are several common information which both pilot projects require.

**Fig. 1** shows a graphic summary of the stakeholders with the corresponding content of the HPCP. The symbol displayed with an arrow to the left represents input provided by stakeholders into the plan. Moreover, the symbol displayed with an arrow to the right shows the information that the stakeholders extract from the plan. Commonalities due to stakeholders and required information with the software function are highlighted in italic. The content is also displayed in italic, even if the stakeholder has not been identified in both release plans. This indicates that the content exists in both plans, but is required by different stakeholders. The respective stakeholder and the respective content that is explicitly required for the ECU or the software function is shown in non-italic.

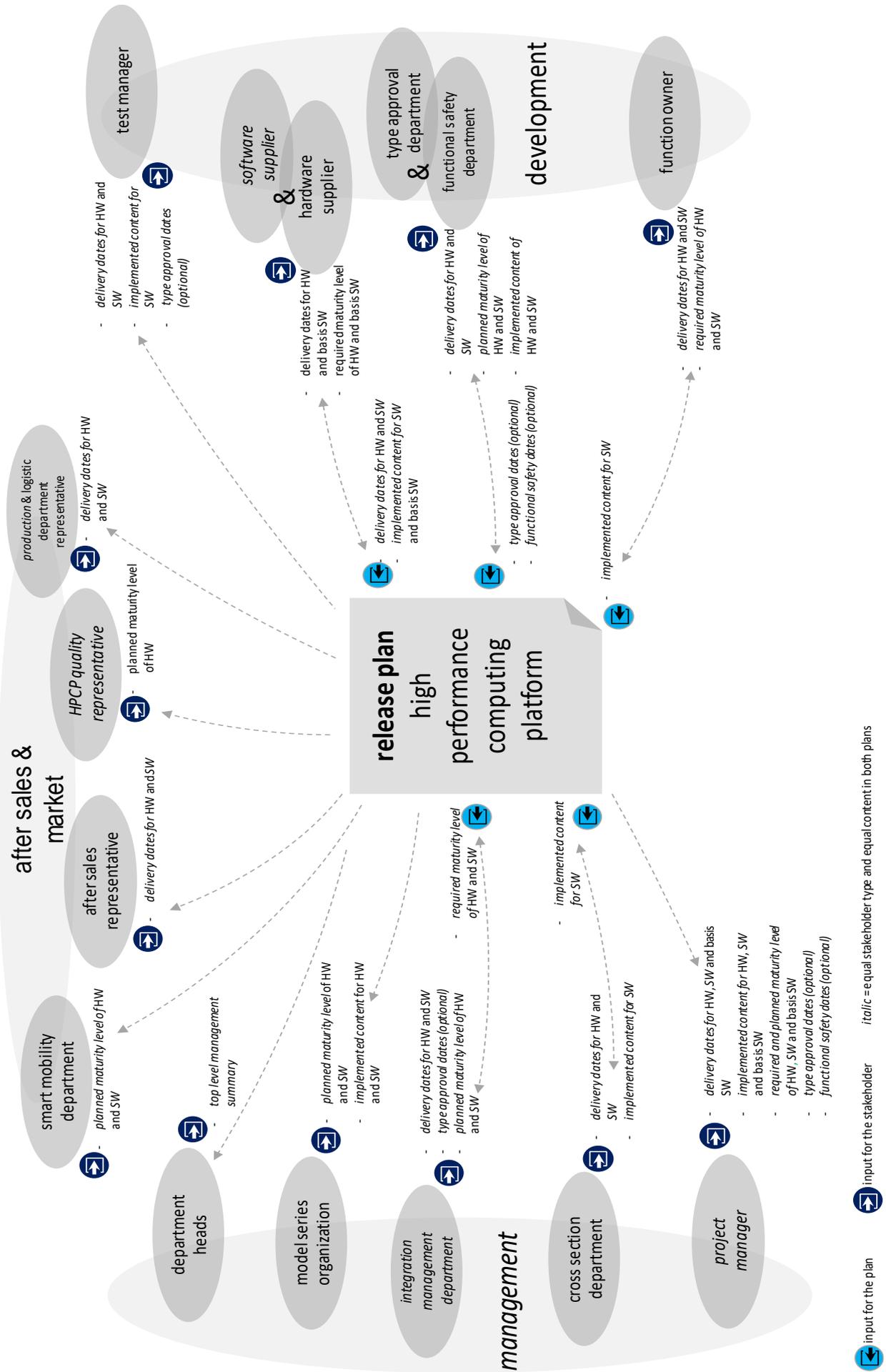

*Fig. 1* Overview of the identified stakeholders with the corresponding information flow of the HPCP

First, we want to take a closer look at the similarities and differences concerning the identified stakeholders (RQ1). The results in Fig. 1 show that not even half of the stakeholders are marked in italic, implying that the release plan of the HPCP requires other stakeholders than the software function. The only shared stakeholders of the release plan are the following six stakeholders: the project manager, the integration management, the HPCP quality representative, the management, production department representative and the respective software suppliers. A possible reason for the few common stakeholders of the plans is the granularity of the planning and the considered planning level of the pilot projects. The release plan of the HPCP maps software and hardware contents on a higher level of abstraction and thus includes the software located on the ECU. Furthermore, it stands out that different categories were formed for the group formation of stakeholders. For example, from a HPCP point of view, a separate category was dedicated to the stakeholder *management*, whereas from a software point of view, *management* belongs to the category *informative stakeholder*. This different classification of stakeholders is summarized in Fig. 2.

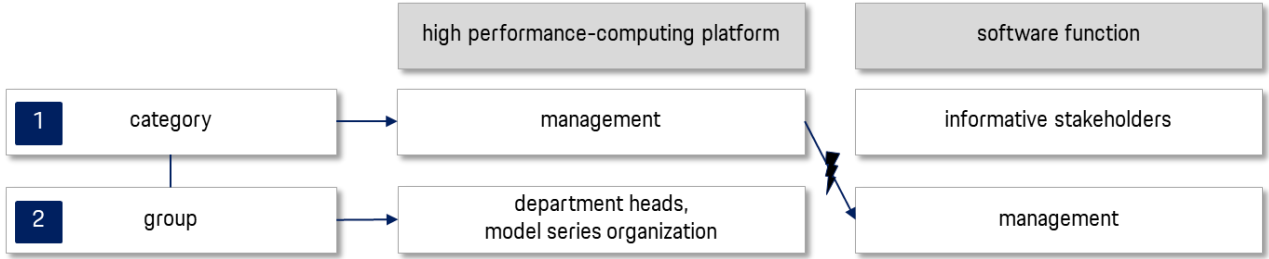

*Fig. 2 different classification of stakeholders*

The identification of stakeholders was executed independently of each other in different research cycles and thus each pilot project has formed its own categories and groups.

In addition, Fig. 1 includes commonalities regarding the contents of the respective plans (RQ2). The italic marked contents are already much more frequent. The strategic framework is not explicitly listed in the graphic as content as it applies to all stakeholders.
Hard constraints such as the required maturity levels given by the company, functional safety dates and type approval dates if necessary are valid for both views. The precise difference regarding the maturity levels results from the particular project-specific adaptation to the project. Contents with reference to the hardware are also only part of the plan of the HPCP and therefore displayed in black.

Fig. 3 illustrates the results of the identified stakeholders and the appropriate information flow from the software function.

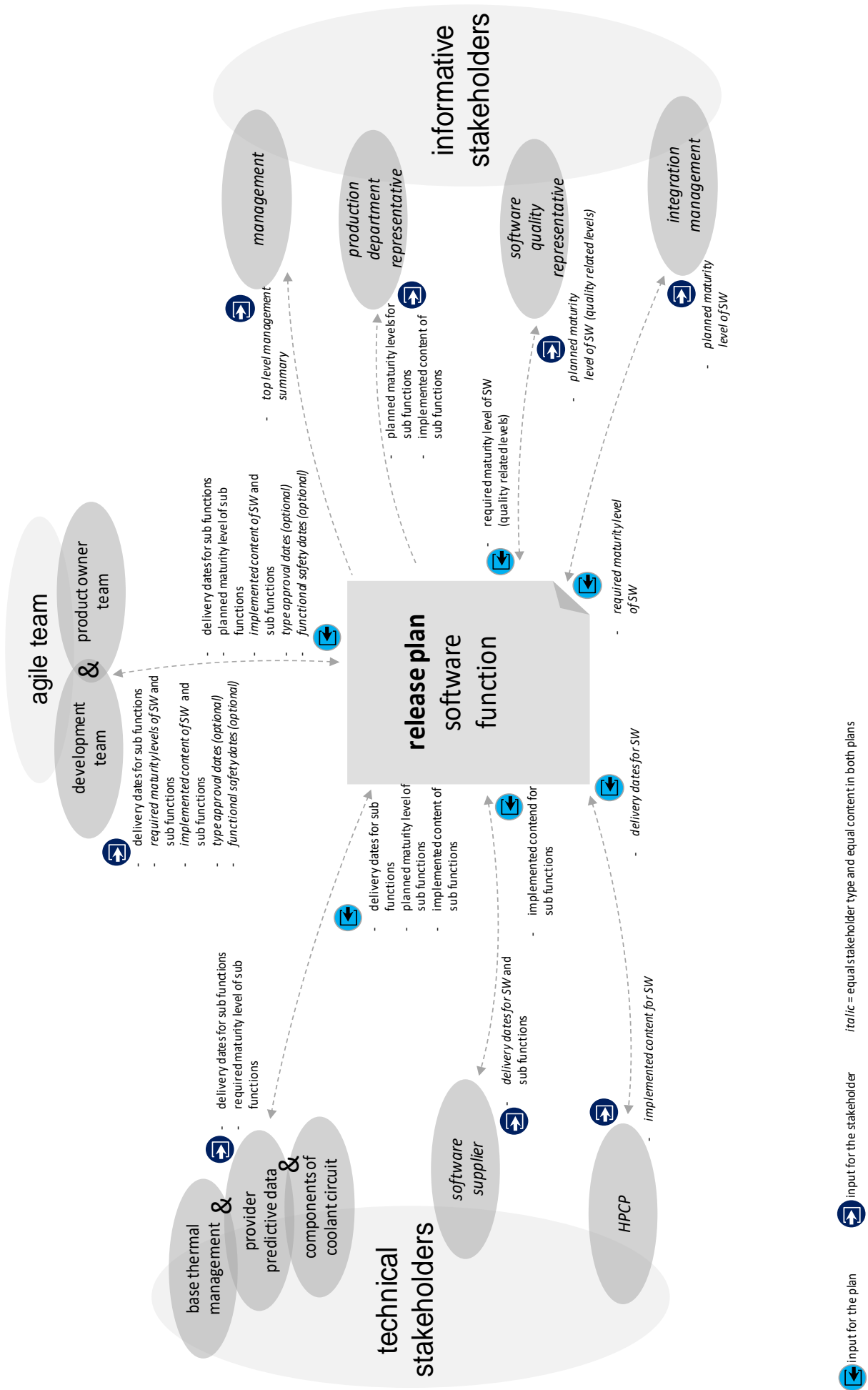

*Fig. 3* Overview of the identified stakeholders with the corresponding information flow of the software function

This Fig. 4 demonstrates that from a functional perspective, fewer stakeholders have been defined. One possible reason for the lower number of stakeholders is that certain stakeholders (for example hardware supplier and logistic department representative) are simply not relevant for release planning from a software planning view. However, the technical interfaces (base thermal management, provider predictive data and components of coolant circuit) are an essential stakeholder for the software function and do not appear in the release plan of the HPCP. Due to the more detailed planning level information related to sub functions are even only part of this plan. The stakeholder test manager is not mentioned separately in the plan of the software function, as he is assigned to the development team.

Both figures demonstrate that only the non-italic marked stakeholders and contents are different and only occur in the respective planning. In summary, the two release plans share similarities and differences, although the release plans focus on different perspectives.
With regard to RQ3, the pursued purpose why a stakeholder needs information from the plan or provides information to the plan, there are no significant differences. Among the identified stakeholders, from both HPCP's and software function's point of view, there are stakeholders who need information for strictly informative purposes or there are stakeholders who need the information for further developing the project.

We conclude this Section 5 by learnings and recommendations for other companies.
Stakeholders will vary in number and role description depending on the company. One of the lessons learned by the results for us is that the number of stakeholders was higher than initially expected. We were astonished to see how many people require release planning. One possible reason for the high number of different stakeholders is the complex development environment with numerous project participants and all kinds of supplier relationships. However, an analysis of the stakeholders is essential for target-oriented release planning if companies want to develop their products effectively and efficiently. By identifying the relevant stakeholders, it is evident who needs something and at what time. This is a valuable guidance for the creation of a release plan, as it allows the OEM's specifications as well as the required legal compliance to be better served. It is displayed preventively and the different stakeholders will not be surprised negatively during the development process. The identification of stakeholders to be considered in release planning allows interfaces to be identified and highlights where dependencies arise. This was a helpful insight for us, as hardware and software development in the automotive industry is a complex process in itself and is in need of any possible support.
We argue that researchers should pay more attention to stakeholders as an influencing factor to release planning und the researchers are supposed to include this factor in their release planning approaches and not simply consider it as given. Practitioners are advised to be more aware of stakeholder requirements. As the results were collected based on two specific projects, the results are not only based on theory but also have a high practical relevance. Real and concrete stakeholders were identified, which could help other companies to identify their stakeholders.

## 6    Conclusion and Future Work

Release planning in a hybrid project environment is increasingly challenging even with new technologies as the HPCP. The different release plans from all involved stakeholders has to be synchronized with co-existing traditional development approaches in the automotive domain. Different stakeholders of two release plans were identified to bring transparency into release planning. In addition, we described the stakeholders who are involved in these release plans and who each have an interest in extracting certain content from it. We outlined the description of this certain content and showed what purpose it serves related to the respective stakeholder. With the presented stakeholders of two release plans, we provide a recommendation for practitioners which stakeholders could be considered due to release planning in the automotive domain. Additionally, we compared the results of both pilot projects and showed the differences and commonalities. The comparison indicated that the two plans have much more in common in relation to stakeholders and content. A possible reason for this might be the complex and non-transparent processes and supplier constellations.

In the future, we will enlarge the results of the two release plans with a more detailed description of dependencies between the stakeholders. Another issue will be the different classification of the stakeholders and it will be analyzed whether a consistent categorization is possible. We will add synchronization points to define what kind of content has to be released with regard to the strategic framework. Furthermore, we will have a closer look at the results if there is a more structured and general procedure possible identifying the stakeholder. For further validation of the stakeholders, we will conduct more workshops with participants from other OEMs or even companies from other regulated domains that are developing complex systems in a hybrid project environment and similar hard constraints. We have demonstrated a deep investigation of two real industry projects considering required stakeholders of a release plan. With the presented examples, we hope to offer support for other OEMs or similar companies that have to struggle with finding the required stakeholders.